\documentclass[12pt,a4paper]{article}
\usepackage[cp1251]{inputenc}
\usepackage{amsmath}
\usepackage{floatflt}
\usepackage{longtable} 
\usepackage{amsthm}
\usepackage[usenames]{color}
\usepackage[dvips]{graphicx}
\usepackage[english]{babel}
\inputencoding{cp1251}
\usepackage{varioref}
\usepackage{amssymb}
\usepackage[plainpages]{hyperref}
\def\bb{\begin{equation}}
\def\ee{\end{equation}}

\def\ve{\varepsilon}

\title{{\bf Autoresonant germ in dissipative system}}

\author{Oleg Kiselev \thanks{Institute of Mathematics, USC RAS({\tt
ok@ufanet.ru})}
       \and
       Sergei Glebov \thanks{Ufa State Petroleum Technical
University({\tt sg@anrb.ru})}          }

\begin{document}

\maketitle

\begin{abstract}
We study an initial stage of autoresonant growth of a solution in a
dissipative system. We construct an asymptotic formula of an autoresonant germ 
that is an attractor for  autoresonant solutions. We present
 a moment of a fall and  a maximum value of the
amplitude for the germ. Numerical simulations
are done.
\end{abstract}

\section{Statement of the problem}

In this paper we study an effect  of a dissipation on a autoresonant  solution. Let us consider the perturbed Duffing's oscillator with a dissipation term:
\begin{equation}
u'' + u + 4\ve^{2/3}\beta u'-2\sqrt{2}u^3 = 4\sqrt{2}\ve f \cos(\omega t).
\label{duffing}
\end{equation}
Here $\ve$ is a small positive parameter, $\omega=(1-\ve^{4/3}t)$,
$\beta>0$ and $f>0$. 
\par
The orders of the dissipation  and perturbation have  special
powers of  $\varepsilon$. It does not lead to a loss of a generality
because these terms contain additional parameters $\beta$ and $f$. On the
other hand it allows us to include these terms in  the primary resonance
equation \cite{BM}.
\par
The autoresonance means the essential  growth of nonlinear oscillations
due to a small oscillating force \cite{Friedland}. This phenomenon appears for (\ref{duffing})
 when the frequency of perturbation decreases slowly and $\beta=0$
\cite{Friedland,Kalyakin,KG}. If $\beta>0$ numerical simulations show that
the growth presents on an initial stage. The direct analysis of the primary resonance equation  allows one to estimate the maximum value of the solution. It was done for a slightly different equation  in \cite{SKKS}. 
\par
We study the solution of (\ref{duffing}) in the form
$$
u\sim \varepsilon^{1/3}\Psi(\tau) e^{i(t-\tau^2)} +\hbox{c. c.},
$$
where $\tau=\varepsilon^{2/3}t$ and c.c. means a complex conjugate term.
The amplitude $\Psi$ of the oscillations is determined by the primary resonance
equation
\begin{equation}
 i\Psi'+(\tau-|\Psi|^2)\Psi+i\beta\Psi=f.
\label{pr}
\end{equation}
\par

\begin{floatingfigure}[r]{8cm} 
\includegraphics[width=8cm,keepaspectratio]{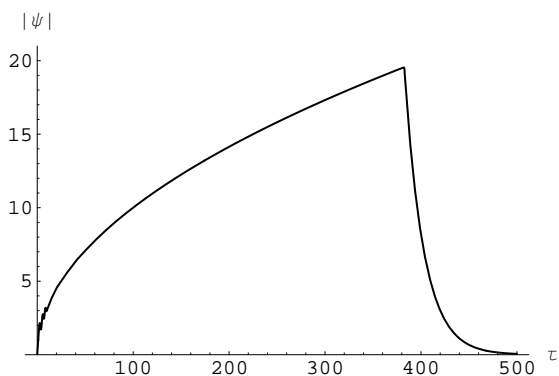}
\caption{\small Modulus of solution for  (\ref{pr}) with $\Psi|_{\tau=0}=0$, $f=1$ and  $\beta=0.05$.}\label{fig1}
\end{floatingfigure}

When $\beta=0$ there exist increasing solutions of (\ref{pr}). These
solutions are related to the autoresonance phenomenon, see
\cite{Friedland,Kalyakin, KG}. 
\par
To clear the problem  we present results of numerical simulations when $\beta>0$.
\par

On the Figure \ref{fig1} one can see the initial stage of growing for
$\Psi$ and the rapid fall. The growing part of the curve is related to the autoresonant phenomenon. The breakpoint shows the boundedness of the autoresonant growing in the system with dissipation.
\par
Our goal is to obtain the  autoresonant germ and study his behavior  up to the moment of fall.
\par

\section{Result}
If $\beta$ is small then  the autoresonant germ has the form
\begin{eqnarray*}
\Psi_G(\tau) \sim \bigg(\sqrt{\tau}+\frac{\sqrt{f^2-\beta^2\tau}}{2\tau}\bigg) \bigg(-\frac{\sqrt{f^2-\beta^2\tau}}{f}-i\beta \frac{\sqrt{\tau}}{f}\bigg)\times \nonumber\\ e^{\displaystyle\frac{i}{2\sqrt{\tau(f^2-\beta^2\tau)}}},\quad  \tau\gg 1. 
\end{eqnarray*}
The life time of this germ is bounded by $\tau_* \sim \displaystyle\frac{f^2}{\beta^2}$ as $\beta\to 0$. This germ is an attractor for captured solutions. The captured solutions approach to the germ as  $O\bigg(\exp\{-\beta\tau\}\bigg)$.
\par

\section{Asymptotic behavior  of autoresonant germ}
\par
When $\beta=0$ there exist pure algebraic solutions of
(\ref{pr}) as $\tau \gg 1$. They were studied in \cite{KG}.  Similar solutions of second order dissipativeless 
equations are called slowly varying equilibriums \cite{Haberman}.
\par
Let us construct a solution of (\ref{pr}) with a slowly varying leading-order term 
 as $\beta>0$. After the following substitution
$$
\theta=\tau\beta^2,\quad \beta\Psi=\varphi
$$
we obtain
\begin{equation}
i\beta^4\varphi'+(\theta-|\varphi|^2)\varphi+i\beta^3\varphi=\beta^3 f.
\label{perturbedPR}
\end{equation}
\par
We construct the  solution of the form
\begin{equation}
\varphi(\theta,\beta)=\bigg(\sqrt{\theta} +\beta^3
\rho_1(\theta)+\beta^4R(\zeta,\theta,\beta)\bigg)\exp\left\{i(\alpha_0(\theta)+\beta\alpha_1(\theta)+\beta^2A(\zeta,\theta,\beta))\right\},
\label{sol1}
\end{equation}
where $0<\beta\ll 1$ and $\zeta=\beta^{-3}\theta$ is a fast variable.
\par
Substitute (\ref{sol1}) into (\ref{perturbedPR}) and gather terms with the same order of $\beta$. It allows us to determine functions $\alpha_0, \alpha_1$ and $\rho_1$ 
\begin{eqnarray*}
\sin(\alpha_0)&=&-\frac{\sqrt{\theta}}{f}, \\
\alpha_1&=&\frac{1}{2\sqrt{\theta(f^2-\theta)}}, \\
\rho_1&=&\frac{\sqrt{f^2-\theta}}{2\theta}.
\end{eqnarray*}
The residual terms $R$ and $A$ are solutions of the system
\begin{eqnarray}
R'_{\zeta}&=&-\beta^2\rho_1'-\beta\rho_1 - \beta^2R+ \beta^{-2}\bigg(-f\sin(\alpha_0+\beta\alpha_1+\beta^2A)+f\sin(\alpha_0)+\nonumber\\
&&\beta f\alpha_1\cos(\alpha_0)\bigg), \label{ResTermsSys}\\
A'_\zeta&=&-\beta\alpha_0'-\beta^2\alpha_1'-\beta^3\rho_1^2-2\beta\sqrt{\theta}R-\beta^5R^2-2\beta^4\rho_1R+\nonumber \\
&f&\hskip-0.35cm\bigg[\cos(\alpha_0+\beta\alpha_1+\beta^2A)-\cos(\alpha_0)\bigg]\left[\left(\sqrt{\theta} +\beta^3
\rho_1(\theta)+\beta^4R\right)^{-1}-\frac{1}{\sqrt{\theta}}\right].\nonumber
\end{eqnarray}
The right hand side of equations contains slowly varying coefficients
with respect to the fast independent variable $\zeta$. Similar equations were
studied by applying WKB-method in \cite{Wasow}. The linearization of (\ref{ResTermsSys}) gives the system with  eigenvalues
$$
\lambda_{1,2} =\pm i(2\theta)^{1/2}\sqrt[4]{f^2-\theta}\beta^{1/2} \mp i\frac{\sqrt[4]{\theta(f^2-\theta)}}{2\sqrt{2}(f^2-\theta)}\beta^{3/2}-\frac{1}{2}\beta^2+O(\beta^{5/2}).
$$
It shows the stability of the solution  with respect to the linear approximation.
  It  means that all
captured solutions $\psi$ are represented by
$$
\psi \sim \bigg(\sqrt{\theta} +\beta^3
\rho_1\bigg)\exp\left\{i(\alpha_0+\beta\alpha_1)\right\} + O(\beta^4\exp\{-\beta^{-1}\theta\}),
\quad \theta < f^2.
$$
As a result we obtain the autoresonant germ 
\bb
\Phi_G(\theta,\beta) \sim \bigg(\sqrt{\theta}+\beta^3\frac{\sqrt{f^2-\theta}}{2\theta}\bigg) \bigg(-\frac{\sqrt{f^2-\theta}}{f}-i \frac{\sqrt{\theta}}{f}\bigg) e^{\displaystyle\frac{i\beta}{2\sqrt{\theta(f^2-\theta)}}}.\label{asymR}
\ee
The representation (\ref{asymR}) allows us to estimate the
 life time of the germ  
$$
\tau_* \sim \left(\frac{f}{\beta}\right)^2. \label{lifeTime}
$$
The maximum amplitude of solutions that  are captured into the autoresonance is 
$$
\max|\Psi| \sim {\frac{|f|}{\beta}}. \label{maxAm}
$$
\par

\newpage

\section{Numerical simulations}

\begin{floatingfigure}[l]{8cm} 
\includegraphics[width=8cm,keepaspectratio]{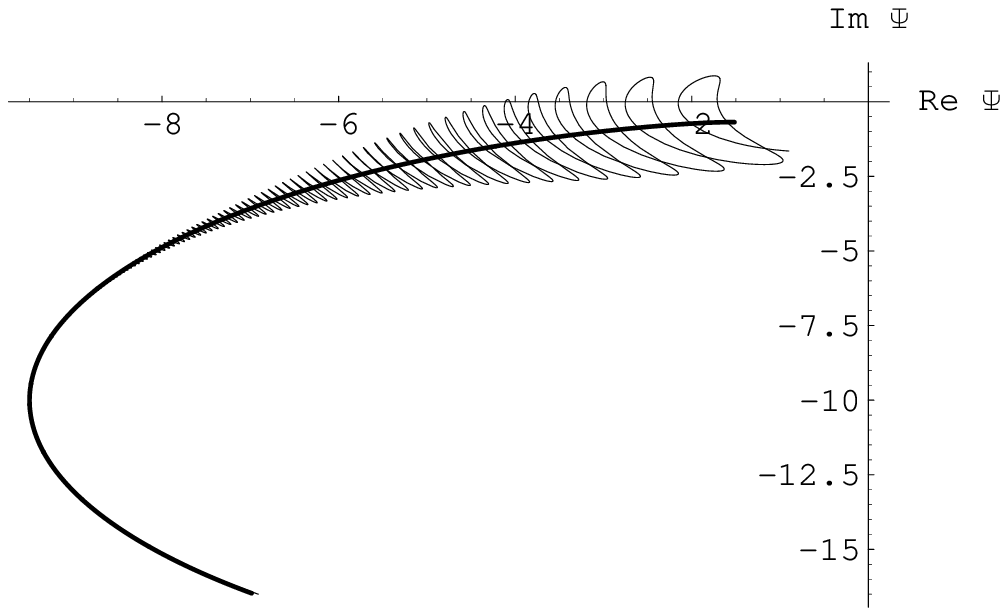}
\caption{\small Sticking of the germ  and numerical solution.}\label{fig2}
\end{floatingfigure}

Here we present the result of  numerical simulations.
Figure \ref{fig2} shows the exponential sticking of the numerical solution and the autoresonant germ.  The heavy line  corresponds to the germ and the thin line shows the behavior  of numerical solution of (\ref{pr}) with  for zero initial data, $f=1$ and $\beta=0.05$.
\par

\vskip1cm

\section{Control of autoresonance with  dissipation}

\begin{floatingfigure}[r]{7cm} 
\includegraphics[width=7cm,keepaspectratio]{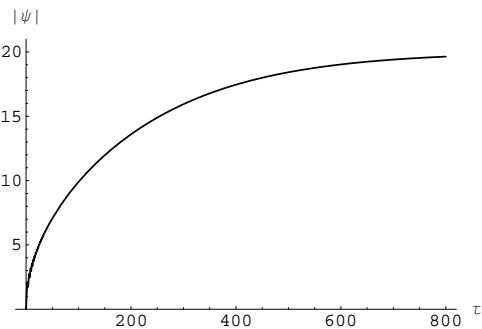}
\caption{\small Control of autoresonant solution of (\ref{pr}) for $\Psi|_{\tau=0}=0$, $f=1$ and
$\beta=0.05$.}\label{fig3}
\end{floatingfigure}

In this section we show a way to control the amplitude of the autoresonant
oscillations in a system with the dissipation. In previous section we demonstrated that solutions fall.  To prevent the fall one should stop the growth of the
frequency of driving force.
\par
Here we present a result of numerical simulations. We suppose to stop the change of
the frequency in equation (\ref{pr}). It leads to a  capture  of the
amplitude. We demonstrate the numerical result when the term $\tau$
in (\ref{pr}) is changed by
$(\beta/f)\tanh\left(f\tau/\beta\right)$ and 
$$
\omega=1-\ve^{2/3}\frac{\beta}{2f\tau}\int^\tau_0\tanh\left(\frac{f\sigma}{\beta}\right)d\sigma.
$$

\section{Conclusions}

In this paper we obtained the autoresonant germ in equation (\ref{duffing})
\begin{eqnarray*}
u\sim \ve^{1/3} \Psi_G(\ve^{2/3}t)\exp\{i(t-\ve^{4/3}t^2) \},\quad t\gg\ve^{-2/3}.
\end{eqnarray*}
We found that
$$
\max |u|\sim \ve^{1/3}\frac{f}{\beta}
$$
and the solution is growing up to the moment $t_* \sim \ve^{-2/3}f^2\beta^{-2}$.
\par
\vskip0.5cm

{\bf Acknowledgments.} This work was supported by grants RFBR  09-01-92436-KE-a                                 and DFG TA 289/4-1 and Grant for Scientific School 2215.2008.1.

\end{document}